\documentclass[12pt,prepint]{aastex}
\usepackage{url}
\usepackage{xcolor}
\usepackage{epstopdf}
\newcommand{\lya}{Ly\ensuremath{\alpha}}

\newcommand{\ha}{H\ensuremath{\alpha}}
\newcommand{\dvlya}{$\Delta v_{Ly\ensuremath{\alpha}}$}
\newcommand{\flamb}{erg~s$^{-1}$~cm$^{-2}$ \AA$^{-1}$~}
\newcommand{\flux}{erg~s$^{-1}$~cm$^{-2}$~}
\newcommand{\oiii}{[O\,{\footnotesize III}]$\lambda$5007}
\newcommand{\oii}{[O\,{\footnotesize II}]$\lambda$3727}
\newcommand{\nv}{N\,{\footnotesize V}$\lambda$1240}
\newcommand{\civ}{C\,{\footnotesize IV}$\lambda$1549}
\newcommand{\siiv}{Si\,{\footnotesize IV}$\lambda$1398}
\newcommand{\kms}{km~s$^{-1}$}

\begin{document}

\title{A z $\sim$ 5.7 \lya\ Emission Line with an Ultra Broad Red Wing}

\author{Huan~Yang\altaffilmark{1}, JunXian~Wang\altaffilmark{1}, Zhen-Ya Zheng\altaffilmark{2}, Sangeeta Malhotra\altaffilmark{2}, James E. Rhoads\altaffilmark{2}, Leopoldo Infante\altaffilmark{3}} 
\altaffiltext{1}{CAS Key Laboratory for Research in Galaxies and Cosmology, Department of Astronomy, University of Science and Technology of China, Hefei, Anhui 230026, China; yanghuan@mail.ustc.edu.cn; jxw@mail.ustc.edu.cn} 
\altaffiltext{2}{School of Earth and Space Exploration, Arizona State University, Tempe, AZ 85287; zzheng13@asu.edu; smalhotr@asu.edu; james.rhoads@asu.edu}
\altaffiltext{3}{Institute of Astrophysics, Ponticia Universidad Catolica, Santiago, Chile; linfante@astro.puc.cl}

\begin{abstract}
Using \lya\ emission line as a tracer of high redshift star forming galaxies, hundreds of \lya\ emission line galaxies (LAEs) at z \textgreater\ 5 have been detected. These LAEs are considered to be low mass young galaxies, critical to the re-ionization of the universe and the metal enrichment of circumgalactic medium (CGM) and intergalactic medium (IGM).
It is assumed that outflows in LAEs can help ionizing photons and \lya\ photons escape out of galaxies.
However we still know little about the outflows in high redshifts LAEs due to observational difficulties, especially at redshift  \textgreater\ 5. Models of  \lya\ radiative transfer predict asymmetric \lya\ line profiles with broad red wing in LAEs with outflows.
Here we report  a z $\sim$ 5.7 \lya\ emission line with a broad red wing extending to $>$ 1000 \kms\ relative to the peak of \lya\ line, which has been detected in only a couple of  z $>$ 5 LAEs till now. 
If the broad red wing is ascribed to gas outflow instead of AGN activity, the outflow velocity could be larger than the escape velocity ($\sim$ 500 \kms) of typical halo mass of z $\sim$ 5.7 LAEs, being consistent with the picture that outflows in LAEs disperse metals to CGM and IGM. 
\end{abstract}


\section{Introduction}
High redshift star forming galaxies can be detected by optical to near-infrared observations of rest-frame ultraviolet light. 
One efficient method selects Lyman break galaxies (LBGs) via the distinctive ÒstepÓ introduced into their blue ultraviolet continuum emission by neutral hydrogen absorption (Steidel et al. 1999; Shapley 2011).  
Another efficient method selects LAEs by their large equivalent width (EW) of \lya\ recombination line due to star formation. 
The standard approach to detect high redshift LAEs has been wide field narrow band imaging survey in gaps of the telluric OH-bands (Cowie \& Hu 1998; Rhoads et al. 2000; Ouchi et al. 2003, 2008; Hu et al. 2004; Wang et al. 2005; Gawiser et al. 2006; Guaita et al. 2010; Tilvi et al. 2010; Nakajima et al. 2012; Cl{\'e}ment et al. 2012; Krug et al. 2012; Shibuya et al. 2012). 
The photometric samples could have contamination rates as large as $\sim$ 40\% (e.g. Wang et al. 2009), and spectroscopy follow up is necessary to remove interlopers (Hu et al. 2004).  
Generally, LAEs are small, low-mass, weakly clustered, young galaxies (Gawiser et al. 2006; Kovac et al. 2007; Pirzkal et al. 2007; Malhotra et al. 2012) and considered to be a subset of LBGs with fainter continuum and smaller dust extinction (Pentericci et al. 2009; Kornei et al. 2010). 

Almost every extreme star forming galaxy has gas outflows. 
Traced by the blue-shifted interstellar metal absorption lines in rest-frame UV or optical spectra, galactic gas outflows are common in local starburst galaxies (Heckman et al. 2000; Veilleux et al. 2005) and ubiquitous in LBGs samples at high redshifts  (Pettini et al. 2001; Adelberger et al. 2003; Shapley et al. 2003; Steidel et al. 2010). 
It is not clear yet whether gas outflows are also prevalent in LAEs, as LAEs samples are biased to galaxies with dim stellar continuum, thus it is difficult to detect the absorption lines. 
Outflows in LAEs and LBGs can regulate star formation in galaxies and are assumed to help ionizing photons escape to ionize the universe and spread metal to circum-galactic medium (CGM) and intergalactic medium (IGM) (eg. Oppenheimer \& Dave 2006).  

Outflows are also key to \lya\ photon escape and to \lya\ emission line profiles. \lya\ is expected to be strong in star forming galaxies.  However, due to the large scattering cross section of \lya\ photons by HI, \lya\ emission could be strongly altered in intensity, kinematics, and apparent spatial distribution (Charlot \& Fall 1993). 
Outflows can help \lya\ photons escape out of local starburst galaxies (Kunth et al. 1998), resulting in P-Cygni profile of \lya\ line (Mas-Hesse et al. 2003).  
In z $\sim$ 2 -- 3 LBGs with \lya\ emission lines, peaks of \lya\ lines are red-shifted relative to galactic systemic redshift while the low ionization interstellar metal absorption lines (tracing gas outflows) are blue-shifted, and the EWs and the velocity offsets of \lya\ lines are closely related to those of low ionization interstellar metal absorption lines (Shapley et al. 2003; Steidel et al. 2010). 
In z $\sim$ 4 -- 6 LBGs sample, peaks of \lya\ emission lines are also red-shifted relative to the low ionization interstellar metal absorption lines (Vanzella et al. 2009; Jones et al. 2012).  
Profiles of \lya\ lines in LBGs are complex, varying from damped absorption to double peak emission, which can be reproduced by models of \lya\ radiative transfer in outflowing gas (Verhamme et al. 2006; Tapken et al. 2007).  
In z $\sim$ 1.8 -- 4.5 gamma-ray burst (GRB) host galaxies with \lya\ emission, the velocity centroid of the \lya\ lines are also redshifted with respect to the galactic systemic velocity, similar to what is seen for LBGs (Milvang-Jensen et al. 2012).

Consistently, in a few z  $\sim$  2 -- 3 LAEs with rest frame optical emission lines (such as [OIII] and H$\alpha$) detected, \lya\ lines are redshifted by $\sim$ 100 -- 300 \kms\ relative to optical lines (McLinden et al. 2011;  Finkelstein et al. 2011; Hashimoto et al. 2013). 
Furthermore,  the composite spectra of eight z  $\sim$  2 -- 3 LAEs (Hashimoto et al. 2013) show blue-shifted interstellar absorption lines relative to optical H$\alpha$ lines. These studies all suggest existence of outflows in LAEs.

However, in most LAEs the \lya\ emission line is the only detectable feature in spectroscopic observations. 
Since the \lya\ emission lines are luminous and their profiles strongly depend on the gas and dust distribution and kinematics, \lya\ lines profiles alone can also be used to trace the interstellar medium properties. 
Hundreds of LAEs at z \textgreater\ 5 have been spectroscopically confirmed by asymmetric \lya\ line profiles that show a sharp blue cutoff (Dawson et al. 2004; Kashikawa et al. 2006, 2011; Sawicki et al. 2008; Hu et al. 2010; Ouchi et al. 2010).  
The asymmetric profile may be caused by absorption and scattering of photons bluer than \lya\ line center by IGM at lower redshift.
However, a few z \textgreater\ 5 LAEs show \lya\ emission lines with very broad red wing, which is ascribed to scattering of \lya\ photons from a shell of gas outflows driven by a powerful starburst (Dawson et al. 2002; Ajiki et al. 2002; Westra et al. 2005). 
Models of \lya\ radiative transfer in galaxies with gas outflows (eg. Verhamme et al. 2006) also predict asymmetric \lya\ profiles with broad red wings. 
We note that in some z \textgreater\ 5 LBGs with \lya\ emission lines detected, the \lya\ line profiles also show very broad red wings (eg. Vanzella et al. 2010; Curtis-Lake et al. 2012).
 
In this work we report  a z = 5.7 \lya\ emission line with a broad red wing located at RA(J2000) = 03:35:39.335, DEC(J2000) = -27:53:15.99 (hereafter J0335) discovered in spectroscopic follow-up of narrow-band selected LAE candidates near the CDF-S field. We discuss the implications of the broad red wing for outflow in LAEs and the potential of using \lya\ as a tracer of galactic outflow at z \textgreater 5. 
We adopt a cosmology with $H_{0} =70$ km~s$^{-1}$~Mpc$^{-1}$, $\Omega_{m} = 0.3$, and $\Omega_{\Lambda} = 0.7$.

\section{Observation and Spectra Analysis Results}
J0335 was selected as a candidate z $\sim$ 5.7 LAE in a deep narrow band imaging survey in a field next to 
the Chandra Deep Field South (CDF-S). We obtained deep NB823 narrowband images using the Mosaic II CCD imager at the Cerro Tololo
Inter-American Observatory (CTIO) 4 m V. M. Blanco telescope on 2005 September 9--11 (UT). The narrowband filter NB823 has a central
wavelength $\lambda_c$ of 823 nm, and a FWHM transmission of 7.5 nm. The broadband B, V and I images used for candidate LAE seletion
are from  ESO Image Survey\footnote{\url{http://www.eso.org/sci/activities/projects/eis/surveys/strategy_DPS.html}} in the same field. 
The optical thumbnail images of J0335 are given in Fig. 1.  
The 5 $\sigma$  limiting Vega magnitudes in a 2.5" aperture of the B, V, I and NB823 images are
25.92,  25.15,   23.83,  and 23.84 mag respectively. 
The galaxy J0335 is only detected in NB823 with a narrow band flux of  3.2$\pm$0.6 $\times$ 10$^{-17}$ \flux. 
As the source is non-detected in the underlying broadband (I band), we use 1$\sigma$ (2$\sigma$)  upper limit of I band flux to estimate the lower limit to equivalent width (EW) of the line. 
Assuming the line is \lya\ line at z $\sim$ 5.7 and simply  following Malhotra \& Rhoads (2002)\footnote{In the calculation, the IGM absorption to the broadband continuum and the Ly$\alpha$ line were ignored since they may cancel each other in the calculation of Ly$\alpha$ EW, see Malhotra \& Rhoads (2002).}, we obtained a lower limit to the rest frame line EW of $>$ 106 \AA\ ($>$ 48 \AA). 
 
The spectroscopic observations of J0335 were taken with the Inamori-Magellan Areal Camera \& Spectrograph (IMACS; Dressler et al. 2006) in multi-slit mode on Magellan I Baade telescope on 2012 October 11. 
The slit width was 1.0$\arcsec$ and the typical seeing during the observation was between 0.5\arcsec\ and 0.7\arcsec. 
The exposures were taken with CTIO-I band filter and 300-l/mm grism with a blaze angle of 26.7 degree, resulting in spectra coverage of 7000\AA-9000\AA\ and a dispersion of 1.25 \AA\ pixel$^{-1}$.  A total of  9000s exposure was obtained. 

The science frames were bias-subtracted, internal flat corrected and wavelength calibrated with the IMACS data reduction package COSMOS\footnote{\url{http://code.obs.carnegiescience.edu/cosmos}}. 
Sky subtractions were performed following the two-dimensional spectroscopy background subtraction method (Kelson 2003).  
We combined the clean 2-d spectra from each individual exposure, and extracted the 1-d spectrum by summing up counts in a 0.8\arcsec\ apertures of the 2-d spectra. Flux calibration was done using the spectroscopic standard star LTT1788. 

The emission line of J0335 clearly showed an asymmetric profile with  a broad red wing in both the 2-d and 1-d spectra (Fig.2).
The asymmetric line profile and non-detections of other emission lines confirm the line is in fact \lya\ emission, but not \oii, \oiii\, or \ha. In particular, the [O\,{\footnotesize II}]$\lambda$$\lambda$3726, 3729 doublet would be resolved at this wavelength with a spectral resolution of $\sim$ 6\AA. 

The observed line flux is (1.71$\pm$0.04)$\times$10$^{-17}$ \flux (measured in a larger 1.6\arcsec\ aperture to avoid aperture loss; the error here represents only statistical uncertainties in the spectrum).
The line luminosity is (6.13$\pm$0.17)$\times$10$^{42}$ erg~s$^{-1}$,  close to $L^*\ $(7-10$\times$10$^{42}$ erg~s$^{-1}$) of the \lya\ luminosity function at z $\sim$ 5.7 (Ouchi et al. 2008; Hu et al. 2010; Kashikawa et al. 2011). 
To estimate spectroscopic line EW, we fit the continuum redward of the \lya\ line (8221.2 - 8800.8 \AA\ in the observed frame)  with a constant, and obtain a continuum flux of 0.021$\pm$0.005 $\times$ 10$^{-18}$ \flamb and a rest frame \lya\ line EW of 121$\pm$29 \AA\ (without correction to IGM absorption to \lya\ line). 

The \lya\ profile showed a marginally resolved narrow line core and a broad red wing. 
The commonly adopted approach to fit high z \lya\ profiles is to use a red continuum plus a half-gaussian profile (setting both components red ward of the line center to zero) convolved with instrument profile. 
For J0335 this approach can fit the narrow line core, but ignores the broad red wing. 
Therefore we add an extra broad Gaussian component to the fitting (Fig.2).
The resulting narrow line core is centered at 8180.5$\pm$0.09 \AA , with FWHM(narrow) = 6.6$\pm$0.3 \AA.  
If we neglect the possible offset of \lya\ line to galaxy systemic redshift, the narrow line center wavelength gives a redshift of 5.7274$\pm$0.0001.  
As the instrument profile resolution is FWHM(instru) $\approx$ 5.9 \AA, estimated by fitting sky line with Gaussian profile, the intrinsic FWHM of the narrow component is $\sim$ 3.0$^{+0.6}_{-0.8}$ \AA\ (108 \kms). 
Since the seeing during the observation (0.5\arcsec -- 0.7\arcsec) was smaller than the slit width (1.0\arcsec), and considering that high redshift LAEs are usually compact in continuum (Malhotra et al. 2012) and in \lya\ emission (Bond et al. 2010), our spectral resolution may be better than 5.9 \AA, and the intrinsic FWHM of the narrow \lya\ core could be $\sim$ 5 -- 6 \AA\ (180 -- 220 \kms). 
The broad component (with signal-to-noise ratio S/N = 5.0) is centered at 8188.4$\pm$2.1 \AA\ with FWHM(broad) = 23$\pm$4 \AA\ (852 \kms). 
Its velocity offset relative to the narrow component is 290$\pm$77 \kms. 


\section{Discussion}

\subsection{AGN or Starburst}
Both AGN and starburst can ionize the surrounding gas and generate \lya\ emission. 
Assuming z = 5.7274,  the expected \nv\ line from an AGN is  sitting on the edge of a strong sky line, and is non-detected. We can only obtain a loose 3$\sigma$ upper limit to \nv/\lya\ $<$ 0.13 (typical value for narrow lines in AGNs is $\sim$ 10\%, Alexandroff et al. 2013). The \siiv\ and \civ\ lines  are out of the spectral range.  The archived mid-IR and X-ray data are too shallow to constrain the AGN activity. 
However, as only $\lesssim$ 5\% of high redshift LAEs are possible AGNs (Malhotra et al. 2003; Wang et al. 2004, 2009; Zheng et al. 2010, 2012, 2013), and J0335 was identified among only a couple of spectroscopically confirmed z $\sim$ 5.7 LAEs in our sample, it is unlikely to be an AGN. 
Although we can not securely rule out AGN activity in J0335 based on current available data, the more likely possibility is that the \lya\ emission line is due to starburst. 
While \lya\ luminosity is admittedly a poor indicator of star formation rate due to radiative transfer effects in galaxies, assuming the \lya\ line is totally due to star formation and taking the standard case B conversion of \lya\ to \ha, we estimate a SFR $\sim$ 6 M$_\sun$/yr (Kennicutt  \& Evans 2012). 

\subsection{Interpretations of \lya\ Profile with Broad Red Wing}
\subsubsection{Outflow Shell Model}
To interpret the \lya\ profile of LBGs and LAEs, previous studies have explored \lya\ emission line profile as a result of a thin shell of outflowing gas driven by starbursts (Verhamme et al. 2006, 2008). 
In their model, gas in far side of outflow can scatter \lya\ photons back to the observer's direction, making photons red-shifted relative to galaxy systemic redshift, avoiding absorption by the material in the near side.  
By changing model parameters such as outflowing velocity, HI column density, velocity dispersion and dust attenuation, a diversity of profiles can be generated. 
The \lya\ line profile (with a broad red wing) of J0335 is qualitatively comparable to those generated by outflow shell model (Verhamme et al. 2006, 2008; Schaerer et al. 2011). 
In particular, an red wing extending to $>$1000 \kms\ can be generated from a very dusty, high column density outflow shell with outflow velocity $\sim$ 300 -- 600 \kms\ (see Fig. 7 of Schaerer et al. 2011).
However, considering the relatively low spectral resolution (FWHM=220\kms) here and that we are unable to determine the model parameters uniquely based on a single line profile, we do not fit the profile with radiative transfer models in this work.  

However, although thin shell outflow model can successfully explain the \lya\ line profiles in many LBGs and LAEs, there are also discrepancies between the outflowing thin shell model and observations. 
Kulas et al. (2012) fitted a sample of z $\sim$ 2 -- 3 LBGs with double peak line profiles and reported clear discrepancies between the models and data. 
Chonis et al. (2013) fitted high resolution \lya\ profiles in three z $\sim$ 2.4 LAEs and also found the model can't fit the profiles well, especially for the two line profiles with a weak and highly blueshifted line peak.\footnote{but see \url{http://www.nordita.org/docs/agenda/slides-alpha2013-schaerer.pdf} for a conference presentation by Daniel Schaerer in September 2013, which gave different results.} 
Furthermore the best fit models usually result in low internal velocity dispersion of the outflowing thin shell in these works. 
This is in contrary to the detected broad interstellar absorption line profiles, which instead  suggest a large bulk velocity range of outflowing gas if we assume that the outflowing shell are of the same material responsible for the interstellar absorption line (Quider et al. 2009; Kulas et al. 2012). 

\subsubsection{Clumpy Outflow at a Large Range of Radii}
Steidel et al. (2010) considered simultaneously the profiles of \lya\ emission and low-ionization interstellar absorption lines in LBGs and suggested a scenario in which the gas outflows are clumpy, spread over a large range in radius and have gradual velocity gradients.  
Photons scattering from the surfaces of discrete clumps would acquire a doppler shift that reflect the velocity of the last scattered clump. 
So the velocity distribution and covering fraction of clumps are most responsible for the kinematics of the observed \lya\ emission line.
To reproduce the profile and velocity offset of the \lya\ line (\dvlya) and the low-ionization interstellar absorption lines in a sample of z $\sim$ 2 -- 3 LBGs, Steidel et al. (2010) constructed a kinematic model where optically thick gas is presented in two kinematic components: one component is at the galaxy systemic redshift and the other is outflowing with a velocity distribution.
The apparent peak of \lya\ emission is modulated primarily by gas at the galaxy systemic redshift.  
When the velocity range spanned by the gas at the galaxy systemic redshift is broader, the \lya\ emission core is more red-shifted (larger \dvlya) and weaker.

Interestingly, an anti-correlation between the EW(\lya) and \dvlya\ (the velocity offset between \lya\ emission and the low-ionization interstellar absorption lines or optical emission lines) has been detected in z $\sim$ 2 -- 3 LBGs and LAEs (Adelberger et al. 2003; Shapley et al. 2003; Hashimoto et al. 2013), supporting the scenario that a smaller \lya\ velocity offset suggests less absorption by gas at the galaxy systemic redshift, thus resulting in a stronger \lya\ line.  
If J0335 is on that trend, its large EW(\lya) implies weak absorption and small \dvlya. 
The relatively low flux of the broad wing compared to the narrow line core suggests that gas clumps with an outflow velocity ranging from zero to larger than 1000 \kms\ has a small sky coverage.

We can compare this outflow velocity with the estimated velocity required for gas to escape the dark matter halo. 
For LAEs at z $\sim$ 5.7 with \lya\ luminosity about $10^{42.6}~erg~s^{-1}$, the average dark matter halo mass is about 6.1$\times$10$^{11}$ M$_\sun$  (Kovac et al. 2007; Ouchi et al. 2010). 
For an isothermal gravitational potential that extends to a maximum radius $r_{max}$, a very rough estimation of the escape velocity at radius r is $v_{esc}(r)=\sqrt{2}~v_{c}~[1+ln(r_{max}/r)]^{\frac{1}{2}}$ (Veilleux et al. 2005). Taking $r_{max}$ = 100 kpc, r = 1 kpc and $v_{c} = \sqrt{\frac{GM_{halo}}{r_{max}}}$, we obtain an escape velocity $v_{esc}$ = 536 \kms. 
As in the equation $r_{max}$ is in square root term and $r_{max}/r$ is in the logarithmic term, changes in the assumed $r_{max}$ and/or r by a of factor 2 would only result in less than 50\% changes in $v_{esc}$. The max outflow velocity in J0335 is larger than the estimated escape velocity, being consistent with the suggestion that low mass galaxies at z \textgreater\ 4 dominate the dissipation of heavy elements into the CGM (Martin et al. 2010) and the enrichment of IGM (Madau et al. 2001; Scannapieco et al. 2002).  

\subsection{Prevalence of Broad Red Wing in \lya\ Profiles of LAEs}
In this subsection we discuss the prevalence of broad red wing in profiles of z $\sim$ 5 -- 6 LAEs by comparing our result with published \lya\ profiles. 
Among the published \lya\ profiles at z = 4.5, 5.7 and 6.5 (Dawson et al. 2007; Hu et al. 2010; Ouchi et al. 2010; Kashikawa et al. 2011), we notice that a few \lya\ line profiles seem to  have broad red wings comparable to our result (eg. 2HC124128+622022 in Fig.A2 of Hu et al. 2010, and J1425554+353039 in Fig.2 of Dawson et al. 2007) but a large fraction of profiles with good S/N do not show broad red wing. 
The reasons why only a small fraction of LAEs show broad red wing may be: 
1) Some LAEs intrinsically lack gas outflows with large velocity, or the covering fraction and/or the column density of outflowing gas is too small,  so the broad red wing is too weak to be detected; 
2) Due to large optical depth of gas at the galaxy systemic redshift, and/or radiative transfer effect in outflowing gas, the apparent peak of \lya\ emission line shifts redward greatly, reducing the significance of broad red wing; 
3) The gas outflow is anisotropic as suggested by simulations (eg. Barnes et al. 2011), so that a broad red wing can only be observed along particular directions. 
By studying \lya\ line profiles with high spectral resolution, good spectral S/N, and other detectable nebulae emission/absorption lines 
(mostly doable for LAEs at lower redshifts, Finkelstein et al. 2011; McLinden et al. 2011; Chonis et al. 2013; Guaita et al. 2013) it is possible to obtain a better understanding of the production of broad \lya\ red wing and the role outflows have in shaping \lya\ line profiles, and enable the use of the \lya\ line as a tracer of gas kinematics at higher redshifts (such as at z \textgreater\ 5).


\begin{figure}[p]
 \begin{center}
  \includegraphics[scale=0.3]{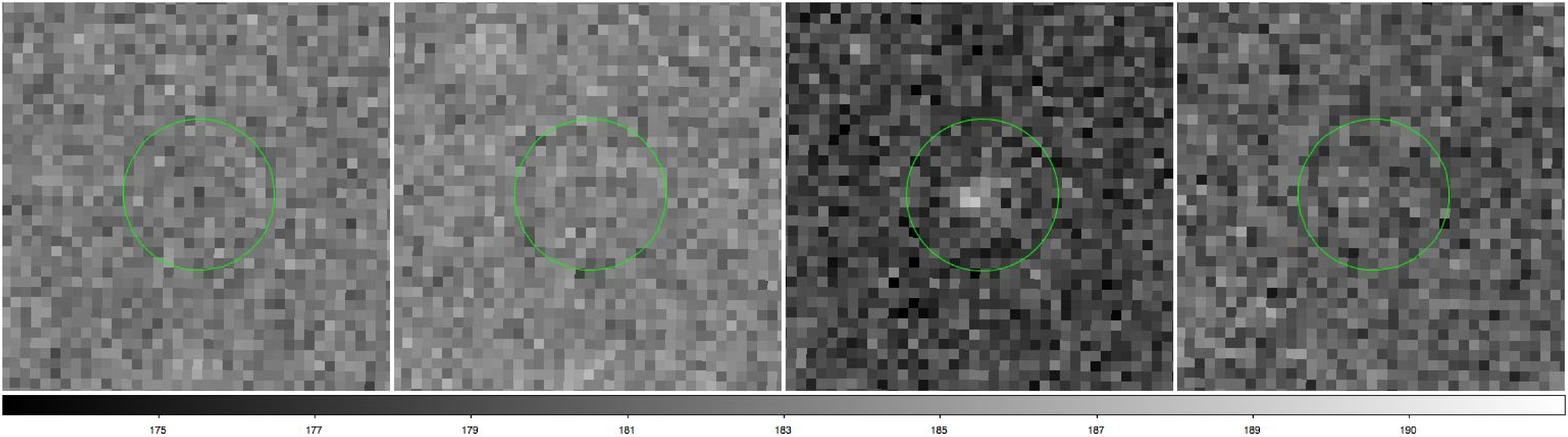}
 \end{center}
 \caption{B, V, NB823 and I band images of J0335 from left to right. The radius of the circles is 2\arcsec. }
\end{figure}

\begin{figure}[p]
 \begin{center}
    \includegraphics[scale=0.8]{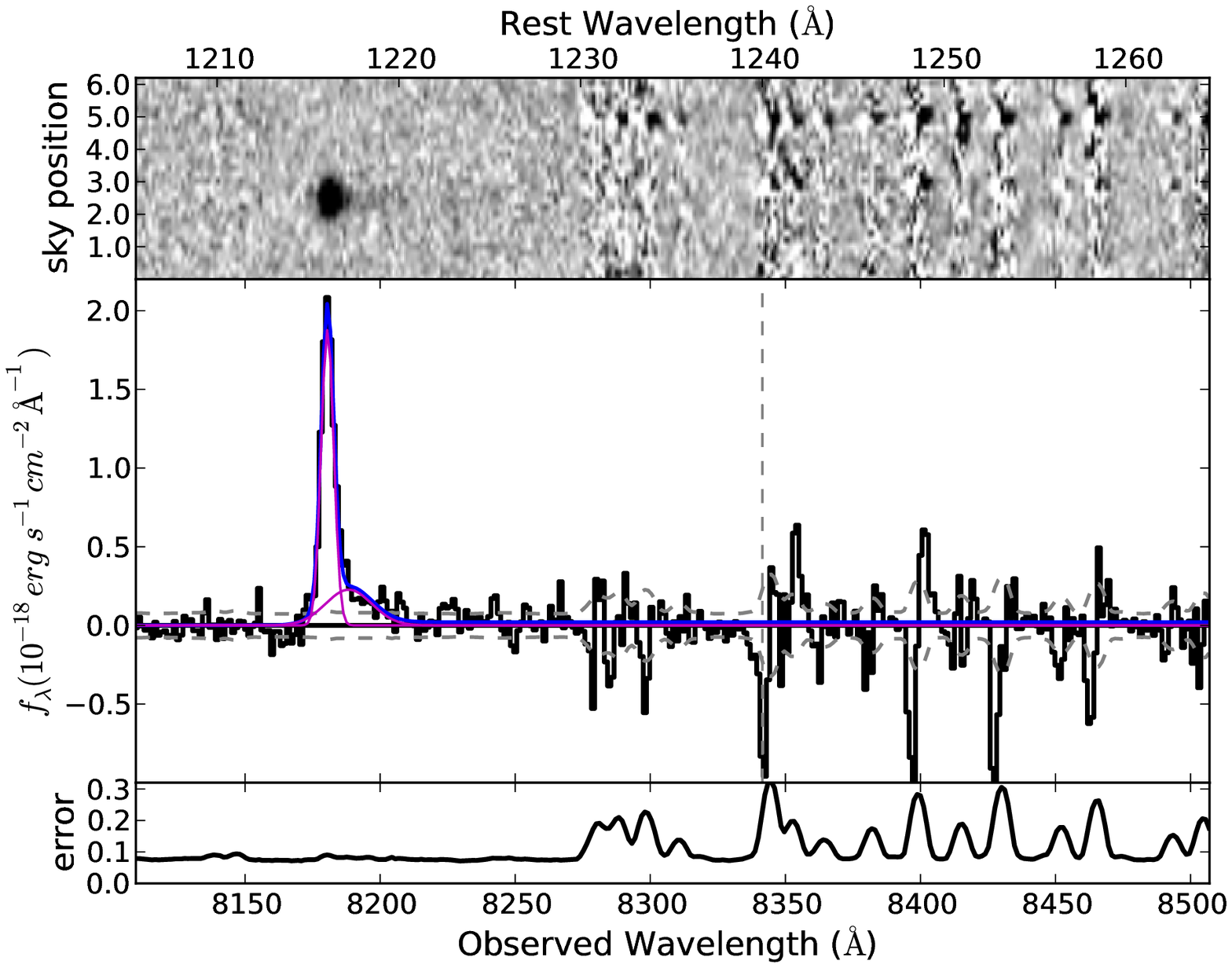}
 \end{center}
 \caption{2-d and 1-d spectra of J0335. 
 Y axis of 2-d spectra is sky position along the slit in units of arcsecond. 
The middle panel show the 1-d spectra (black), best fit model (blue solid line), two Gaussian components (magenta solid lines) and $\pm$1$\sigma$ error of the spectrum (grey dashed line, 
also plotted in the bottom panel. The position of \nv\ line is marked with vertical dashed line.).    
Both fluxes and errors of 1-d spectra are in unit of 10$^{-18}$ \flamb.
}
\end{figure}

\acknowledgments
We acknowledge support from National Natural Science Foundation of China through grant 10825312 \& 11233002.
JXW acknowledges support from Chinese Top-notch Young Talents Program.
This research uses data obtained through the Telescope Access Program (TAP), which is funded by the National Astronomical Observatories, Chinese Academy of Sciences, and the Special Fund for Astronomy from the Ministry of Finance.
We would like to thank the scientists and telescope operators at Magellan telescope for their help. LI acknowledges support from CONICYT  CATA Based program.


\end{document}